\documentclass{article}
\usepackage[utf8]{inputenc}
\usepackage{pdfpages}
\usepackage[a4paper, total={6in, 8in}]{geometry}
\usepackage{setspace} 
\title{Bioinformatics: a working concept}
\author{Ben Hesper, Paulien Hogeweg}

\begin{document}

\large 
{{BIO-INFORMATICS: a working concept.\\ A translation of ``Bio-informatica: een werkconcept'' by B. Hesper and P. Hogeweg}}

\singlespacing 
{\fontfamily{qtm}\selectfont
\small 
Translated by Jeroen Meijer (November 2021).\\
Followed by the original article in Dutch which originally appeared as:\\
Hesper, B and Hogeweg, P (1970) ``BIO-INFORMATICA: een werkconcept''. \emph{Het Kameleon} 1.6, pp. 28-29.
}
\onehalfspacing 

\vspace{12pt}

\normalsize

\underline{}

\vspace{8pt}
What kind of theory we gain from the scientific study of systems strongly depends on our choice of model. 
By `model' we here mean a representation of the aspects that currently interest us of the studied object, in a form that is tractable for analysis. 
For example: when we are interested in the interactions between genetic factors for mutations, but at the moment do not care about the molecular structure of genes, we find that representing genes as points on a line that represents a chromosome allows analytical handling (interpretation, prediction). 

This, then, is a model: the `factorgenetic’ model. If we choose a mathematical-stochastic representation we will get the population genetic model, if we choose a representation in terms of chemical structures we will get a molecular genetic model. (Note that all 3 models (not just the mathematical model) are exact, that is to say, lead to unequivocal statements.) 

In the study of dynamical systems we can distinguish 2 classes of models: one that represents the system as an energy processing system, and one that represents the system as an information processing system. For example: if we study the nervous system in the light of metabolism, oxygen uptake etc., then we speak in terms of energy, chemical potential, etc.; if we study it in the light of action potentials, sensory and motor control etc., then we use terminology such as information, pulse patterns, etc.

For this reason it is often useful in physical and biological research to distinguish between studying the energetic phenomena and studying the ‘informatic’ phenomena in the studied object. For this classification one generally uses the collective terms: ‘energetics’ and ‘informatics’. Note that currently an important subfield of informatics is the academic field of using computers for information processing. The term ‘informatics’ is often used specifically for this subfield. This is of course no problem, as long as we keep in mind that a broader interpretation is possible. 

Within informatics we can distinguish subfields based on the physical phenomenon that ‘carries’ the information, e.g. electronics (flow of electrons as information carriers), fluidics (flow of fluids), photonics (flow of photons), neuronics (dynamics of neural action potentials), similar to how we distinguish mechanics, kinetics, etc. as subfields in energetics. 


We can also classify scientific disciplines based on the type of object we work with, namely with a prefix. We do this for example when talking about geophysics, hydrodynamics, biomechanics, biomathematics. 

As information processing plays a very important role in living systems (they are, after all, needed to maintain ‘order’ and are therefore at least as important as energetic processes), it is useful to speak about ‘bio-informatics’. And so with this we mean: the study of information processing in biological systems. Examples: the functioning of the nervous system, hormonal regulation, interactions in ecosystems, genetic information transfer. 

We should distinguish this from the application of informatics in biological research, e.g. the use of computers, electronics, photographic methods. Because this is just informatics and no different from informatics applied in other fields.

\vspace{20pt}
\hspace{270pt} Ben Hesper

\hspace{270pt} Paulien Hogeweg
\newpage
\thispagestyle{empty}
\includegraphics[scale=0.85,page=1]{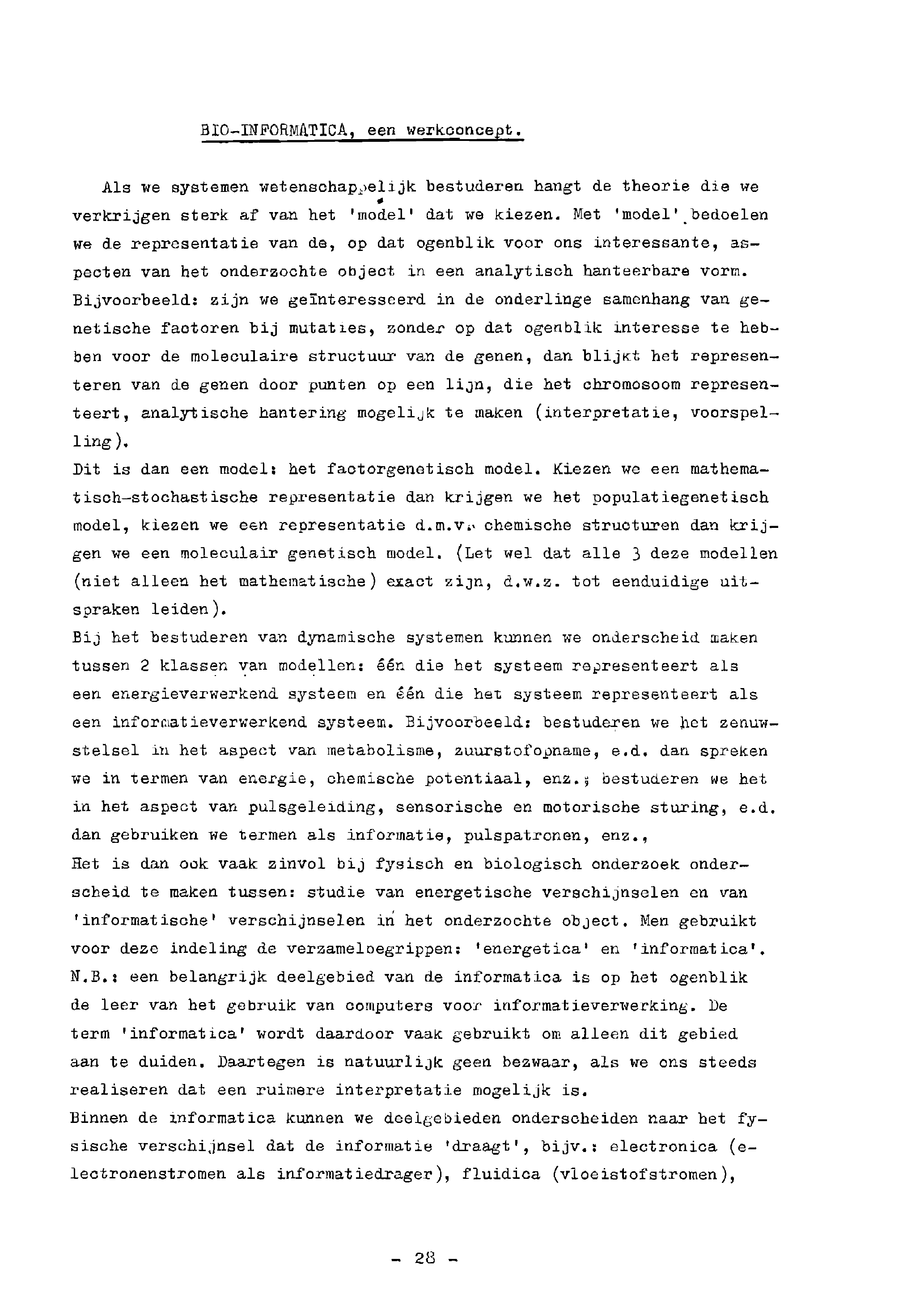}
\newpage
\thispagestyle{empty}
\includegraphics[scale=0.85,page=2]{HH70.pdf}
\end{document}